\documentclass[%
reprint,
 groupedaddress,
showpacs,
 amsmath,amssymb,
 aps,
prl,
]{revtex4-1}

\usepackage{graphicx}
\usepackage{dcolumn}
\usepackage{bm}

\begin{document}
\preprint{Draft 5}

\title{
Multiple Diffusion--Freezing Mechanisms in Molecular Hydrogen Films}

\author{T. Makiuchi}
\altaffiliation[Current address: ]{Department of Applied Physics, The University of Tokyo, Bunkyo 113-8656, Japan.}
 \author{K. Yamashita}
 \author{M. Tagai}
\author{Y. Nago}
\author{K. Shirahama}
\email{keiya@phys.keio.ac.jp}
\affiliation{Department of Physics, Keio University, Yokohama 223-8522, Japan}

\date{\today}

\begin{abstract}


Molecular hydrogen is a fascinating candidate for quantum fluid showing bosonic and fermionic superfluidity.  
We have studied diffusion dynamics of thin films of H$_2$, HD and D$_2$ adsorbed on a glass substrate by measurements of elasticity.  
The elasticity shows multiple anomalies well below bulk triple point. 
They are attributed to three different diffusion mechanisms of admolecules and their ``freezing'' into localized state: classical thermal diffusion of vacancies, quantum tunneling of vacancies, and diffusion of molecules in the uppermost surface. 
The surface diffusion is active down to 1 K, below which the molecules become localized. 
This suggests that the surface layer of hydrogen films is on the verge of quantum phase transition to superfluid state. 

 \end{abstract}

\maketitle

Light molecules such as hydrogen\cite{Silvera1980, VanKranendonkBook} and helium form quantum fluids and solids, in which quantum effects emerge. 
If hydrogen molecules are kept delocalized at low temperatures, 
exchange between molecules 
bring about quantum effects. 
Among the quantum effects, superfluidity in liquid phase (and even in solid, so called supersolidity) is an extraordinary but a fundamental phenomenon caused by macroscopic quantum coherence. 
Although study of superfluid helium has spanned almost a century, novel superfluids such as ultracold atoms\cite{Anderson1995}, and condensates of polaritons\cite{Kasprzak2006} and magnons\cite{Borovik-Romanov1984,Demokritov2006} 
have innovated research fields of condensed matter physics. 
Molecular hydrogen can be unique superfluids\cite{Ginzburg1972,Maris1983}: 
Two nuclear spin isomers in bosonic H$_2$ and D$_2$, namely ortho and para states, produce spin--dependent Bose--Einstein condensates. 
HD, the fermionic isotope, is even more interesting as it may produce anisotropic 
superfluids in which Cooper pairs possess unprecedented internal degrees of freedom.

In contrast to helium, bulk liquid hydrogen (e.g. H$_2$) solidifies below 13.8 K and shows no superfluidity because of stronger 
attraction than that of helium. 
Efforts of realizing superfluidity in H$_2$ have therefore been concentrated to weaken 
the intermolecular attractive forces by reducing dimension or system size \cite{Gordillo1997,Khairallah2007,Mezzacapo2008}. 
The only indication of superfluidity was found in an experiment of nanoclusters with about fifteen H$_2$ molecules\cite{Grebenev2000}: They cannot be regarded as a macroscopic quantum effect. 
Experiments to search for superfluidity in 
H$_2$ films on solid surfaces and H$_2$ in confined geometries were unsuccessful\cite{Vilches1992,Torii1990}, and simulations 
are controversial on the existence of superfluid phase\cite{Gordillo1997,Dusseault2018}.

In this work, we have studied the dynamics of thin hydrogen films by a new technique of elasticity measurement. 
The elastic study is motivated by our recent finding in helium and neon films\cite{Makiuchi2018,Makiuchi2019}. 
$^4$He films show superfluidity when the coverage $n$ exceeds a critical value $n_{\mathrm c}$ ($\sim 20\ \mathrm{\mu mol/m^2}$, roughly 1.8 layers)\cite{Makiuchi2018}. 
The emergence of superfluidity occur as a quantum phase transition (QPT) between localized solid and superfluid at $n = n_{\mathrm c}$. 
We have found that $^4$He, $^3$He, and $^{20}$Ne films show an ``elastic anomaly'', in which the elastic constant of localized films measured with AC strain increases at low temperatures with an excess dissipation. 
In $^4$He and $^3$He films, the characteristic temperature of the stiffening decreases as $n$ approaches $n_{\mathrm c}$. 
The elastic anomaly is quantitatively explained by thermal activation of helium atoms from the localized states to mobile, extended states with energy gap $\Delta$.
The gap $\Delta$ decreases to zero obeying a power law $\Delta \propto (n - n_{\mathrm c})^{\alpha}$ with $\alpha \sim 1.3$ (1.8) for $^4$He ($^3$He).
Therefore, the critical coverage $n_{\mathrm c}$ is identified as a coverage at which the elastic anomaly disappears by gap closure, and (super)fluidity emerges as helium atoms occupy the extended states. 
On the other hand, in neon film, similar elastic anomaly is observed but the characteristic temperature does not decrease below 5 K: The energy gap does not close and neon film does not show QPT.

These results suggest that elastic anomaly can examine the existence of QPT and superfluidity in adsorbed molecules. 
In this Letter, we apply this idea to films of three hydrogen isotopes, H$_2$, D$_2$, and HD. 
We have found multiple elastic anomalies in hydrogen films, unlike the single elastic anomaly in helium and neon films. 
Each elastic anomaly 
corresponds to a ``freezing'' of diffusive motion of hydrogen molecules. 
They are identified as classical thermal diffusion, quantum tunneling, and surface diffusion.
Although no 
QPT was observed, the uppermost surface layer of hydrogen films is on the verge of QPT to superfluid state.


\begin{figure*}
 \centering
 \includegraphics[width=\textwidth]{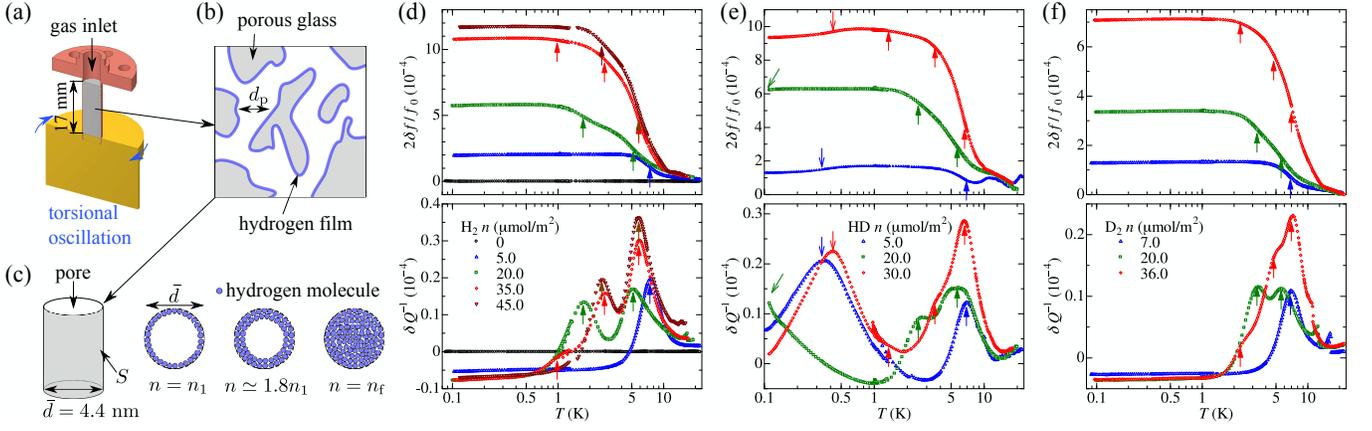}
 \caption{
 (a) Cross-section of the torsional oscillator.
 (b) Schematic cross-section of the porous glass.
 (c) Cylindrical representation of the pore, and sketches of the monolayer, two-layer and full-pore coverages.
 (d--f) Temperature dependence of the normalized resonant frequency shift and the excess dissipation of (d) H$_2$, (e) HD, and (f) D$_2$.
 The upward arrows indicate peaks of $\delta Q^{-1}$ and locally maximal slopes in $2\delta f/f_0$.
 The data of $2\delta f/f_0$ in (e) and (f) are vertically shifted to set $2\delta f/f_0= 0$ at high $T$.
 Only in HD data (e), a dissipation peak accompanied with decrease in $2\delta f/f_0$ is indicated by downward arrows (see 
text).
 }
 \label{fig:Fig1}
\end{figure*}

The elastic measurement was carried out with the same torsional oscillator (TO) used in the previous helium and neon studies\cite{Makiuchi2018,Makiuchi2019}.
The 
TO consists of a cylindrical BeCu torsion rod embedded with a porous glass rod and a metal bob [see Fig. \ref{fig:Fig1}(a)].
The porous glass called Gelsil has three-dimensionally connected nanopores [see Fig. \ref{fig:Fig1}(b)].
From a N$_2$ adsorption isotherm, we obtained the surface area $S=166\ \mathrm{m^2}$, the total pore volume $v_\mathrm{p}=0.184$ cm$^3$, the peak value of the pore diameter $d_\mathrm{p}=$ 3.9 nm, and the porosity $p=0.54$.
If we consider the nanopore a cylindrical pore with uniform diameter, the mean diameter is $\bar{d}=4.4$ nm, which is about 11 times larger than the diameter of a hydrogen molecule.
The TO 
was mounted on a plate thermally linked to the mixing chamber of a dilution refrigerator.

 The coverage $n$ is the amount of dosed molecules divided by $S$. 
The coverage at which a monolayer is formed [see Fig. \ref{fig:Fig1}(c)] is estimated to be $n_1 = (v_\mathrm{m}^2 N_\mathrm{A})^{-1/3}$, where $N_\mathrm{A}$ is the Avogadro constant and $v_\mathrm{m}$ 
the molar volume. 
Using $v_\mathrm{m}=$ 23.30, 21.84, and 20.58 $\mathrm{cm^3/mol}$ \cite{Roder1973}, we get $n_1=$ 14.5, 15.2, and 15.8 $\mathrm{\mu mol/m^2}$ for H$_2$, HD, and D$_2$, respectively.
The amount of molecules at which the pore is fully filled (the full-pore) is $n_\mathrm{f} = v_\mathrm{p}/Sv_\mathrm{m}$, and we have $n_\mathrm{f}=$ 47.6, 50.8, and 53.9 $\mathrm{\mu mol/m^2}$ for H$_2$, HD, and D$_2$, respectively.
To avoid solidification, 
hydrogen gas\cite{samplegas} was introduced through a capillary which was thermally isolated from cold stages.
The film was then annealed above the triple point temperature $T_{\mathrm{tp}}$ (13.8, 16.6, and 18.7 K for H$_2$, HD, and D$_2$, respectively) followed by slow cooling down to 0.1 K.
The data was taken during a warming from 0.1 to 1.3 K with normal operation of dilution refrigerator, and a subsequent warming from 1.0 to 22 K without circulating $^3$He. 
Data at 1.0--1.3 K were doubly measured due to this procedure. 
No critical effect was observed from the annealing conditions and possible ortho--para conversion. 

The resonant frequency $f$ and energy dissipation $Q^{-1}$ of torsional oscillation represents the elastic constant and energy loss of the substrate-hydrogen composite system.
We refer to the frequency and dissipation without hydrogen film ($n=0$), $f_\mathrm{B}(T)$ and $Q^{-1}_\mathrm{B}(T)$, as the background.
When the hydrogen film is formed on the pore surface, $f$ and $Q^{-1}$ change by the elastic contribution of the film.
The elastic constant and dissipation of the hydrogen film are given by a normalized 
frequency shift $2\delta f/f_0$ and an excess 
dissipation $\delta Q^{-1}$, where
\begin{align}
 \delta f(T) &= f(T) - f_\mathrm{B}(T), \label{eq:df} \\
 \delta Q^{-1}(T) &= Q^{-1}(T) - Q^{-1}_\mathrm{B}(T). \label{eq:dQ}
\end{align}

Figure \ref{fig:Fig1}(d) shows $2\delta f/f_0$ and $\delta Q^{-1}$ of H$_2$ film as a function of $T$.
At a small coverage of $n=5.0\ \mathrm{\mu mol/m^2}$ ($\sim 0.3n_1$), $2\delta f/f_0$ increases toward low $T$ with a single peak of $\delta Q^{-1}$ at $T_\mathrm{p} =7.5$ K.
$2\delta f/f_0$ has the largest slope at the dissipation peak temperature $T_\mathrm{p}$. 
The negative value of $\delta Q^{-1}$ at low $T$ indicates that 
the hydrogen film reduces the internal loss of the porous glass.
These behaviors of the elastic anomaly are qualitatively similar to those observed in the helium and neon films \cite{Makiuchi2018,Makiuchi2019}. 
However, at coverages of 20.0, 35.0, and 45.0 $\mathrm{\mu mol/m^2}$, $\delta Q^{-1}$ has multiple peaks, unlike in the helium and neon cases.
The number of peaks are two at 20.0 and 45.0 $\mathrm{\mu mol/m^2}$, and three at 35.0 $\mathrm{\mu mol/m^2}$.
At $n=45.0\ \mathrm{\mu mol/m^2}(\sim n_\mathrm{f})$, the pore is almost filled with H$_2$.
For every coverage, the high-$T$ value of $2\delta f/f_0$ almost equals the background value, meaning that the hydrogen film is soft (even below 
$T_{\mathrm{tp}}$), 
while the low-$T$ value of $2\delta f/f_0$ has a $T$--independent increment, which means that the film is stiff. 
The shear modulus given by a formula \cite{Makiuchi2019} $G=(G_\mathrm{g0}/0.197)(2\delta f/f_0)/(1-p)^2$ is 190 MPa at $45.0\ \mathrm{\mu mol/m^2}$ at low-$T$. 
This is the same order of magnitude as $G=100$ MPa of solid hydrogen \cite{Silvera1980}.

The multiple elastic anomalies were also observed in HD and D$_2$ films. 
This is shown in Fig. \ref{fig:Fig1}(e) and \ref{fig:Fig1}(f). 
In the HD and D$_2$ experiments, $2\delta f/f_0$ sometimes showed unexpected shifts in the entire $T$ range after changing $n$.
We attribute these $T$--independent shifts to an effect of vibrational disturbance. 
We set $2\delta f/f_0=0$ at high $T$ by vertically shifting the data. 
In HD films, we observed fourth dissipation peak accompanied with a slight decrease of $2\delta f/f_0$ below about 1 K, i.e., the HD film is slightly softened at lowest temperatures. 
This anomaly was seen only in HD, and the origin is unknown. 
We will not discuss this fourth anomaly in this Letter.

In the previous helium and neon studies \cite{Makiuchi2018,Makiuchi2019}, it was established that the elastic anomaly originates from anelastic relaxation process with thermal activation of molecules \cite{NowickBook,Tait1979}. 
The relation between the activation energy (the energy gap), $E$, and the thermal relaxation time, $\tau$, is $\tau=\tau_0 \exp(E/k_\mathrm{B}T)$, where $\tau_0^{-1}$ is the attempt frequency.
When $\omega \tau \gg 1$ (low $T$) with $\omega = 2\pi f$, the molecules are localized, thus $\delta f>0$. 
On the other hand, when $\omega \tau \ll 1$ (high $T$), the molecules are frequently activated to mobile states during deformation of the substrate, therefore $\delta f$ is essentially zero.
The crossover occurs at $T_\mathrm{p}$ at which $\delta Q^{-1}$ has a peak and $\omega \tau(T_\mathrm{p}) =1$.

\begin{figure}
 \centering
 \includegraphics[width=65mm]{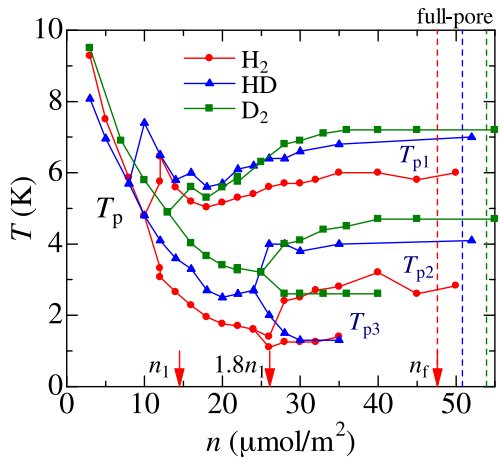}
 \caption{The coverage dependence of the dissipation peak temperature $T_\mathrm{p}$. 
 The monolayer ($n_1$), the two-layer ($1.8n_1$), and the full-pore ($n_\mathrm{f}$) coverages of H$_2$ are indicated by arrows.}
 \label{fig:Fig2}
\end{figure}

The multiple dissipation peaks 
indicate that the hydrogen film freezes by different mechanisms that have unique activation energies.
The coverage dependencies of $T_\mathrm{p}$'s are displayed in Fig. \ref{fig:Fig2}.
The overall features of $T_\mathrm{p}$ are similar among H$_2$, HD, and D$_2$.
The curve of $T_\mathrm{p}$ splits twice into branches.
We label the peak temperatures at the branches in descending order, i.e., $T_\mathrm{p1}>T_\mathrm{p2}>T_\mathrm{p3}$, as indicated in Fig. \ref{fig:Fig2}. 
The curve of $T_\mathrm{p1}$ appears at a coverage less than $n_1$, and the curve of $T_\mathrm{2}$ appears at the two-layer coverage $n_2\simeq 1.8 n_1$.
The curves of $T_\mathrm{p1}$ and $T_\mathrm{p2}$ persist until $n>n_\mathrm{f}$, but the dissipation peak corresponding to the curve of $T_\mathrm{p3}$ disappears at about 40 $\mathrm{\mu mol/m^2}$. 
This is shown as the termination of the $T_\mathrm{p3}$ curve. 
From these behaviors, we conclude that $T_\mathrm{p1}$ and $T_\mathrm{p2}$ are related to activation energies in solid hydrogen, and $T_\mathrm{p3}$ originates from activation in the surface of films. 
It is remarkable that these diffusion mechanisms are indistinguishable below about 10 and 25 $\mathrm{\mu mol/m^2}$ where two branches of $T_\mathrm{p}$ merge.

$T_\mathrm{p1}$ shows a shallow minimum at about 18 $\mathrm{\mu mol/m^2}$ and stays constant (5--7 K) at higher coverages. 
The coverage dependence of $T_\mathrm{p1}$ is \textit{quantitatively} similar to that of $T_\mathrm{p}$ of neon film \cite{Makiuchi2019}.
This is explained by the fact that the Lennard--Jones potentials for hydrogen and neon are almost identical \cite{Nosanow1977}. 
Therefore, the first anomaly around $T_\mathrm{p1}$ is caused by classical molecular interaction.

The ratio $T_\mathrm{p1}/T_\mathrm{p2}\sim 2$ is common to 
all isotopes.
In bulk solid hydrogen, 
 two self-diffusion mechanisms are expected \cite{Ebner1972}.
One is the classical thermal diffusion of vacancies in solid, and the other is the quantum tunneling of vacancies.
The classical 
diffusion is related to $T_\mathrm{p1}$. 
The activation energies for the thermal diffusion and the quantum tunneling are $E_\mathrm{v}+E_\mathrm{b}$ and $E_\mathrm{v}$, respectively, where $E_\mathrm{v}$ is the vacancy formation energy and $E_\mathrm{b}$ is the potential barrier.
As the ratio holds $(E_\mathrm{v}+E_\mathrm{b})/E_\mathrm{v}\sim 2$ for solid H$_2$ and D$_2$ \cite{Ebner1972}, we conclude that $T_\mathrm{p2}$ of our system is related to the quantum tunneling.

$T_\mathrm{p3}$ vanishes as $n$ approaches $n_\mathrm{f}$. 
This is due to the disappearance of the surface of the film as the film thickens inside nanopores [see Fig. \ref{fig:Fig1}(c)]. 
Therefore, the third anomaly $T_\mathrm{p3}$ is attributed to the diffusion in the surface of the 
 film. 
It has been suggested that the surface layer of hydrogen films is mobile 
at low temperatures \cite{Maruyama1993,Sukhatme1996,Bloss2000}. 
The surface diffusion at 
low temperatures is because the motion of molecules at the uppermost layer does not require formation of vacancy \cite{Sukhatme1996}.
The elastic anomaly at $T_\mathrm{p3}$ clearly indicate the existence of the surface fluid state in all the isotope 
films down to 1--2 K, a temperature of one tenth of the triple point.

\begin{figure}
 \centering
 \includegraphics[width=60mm]{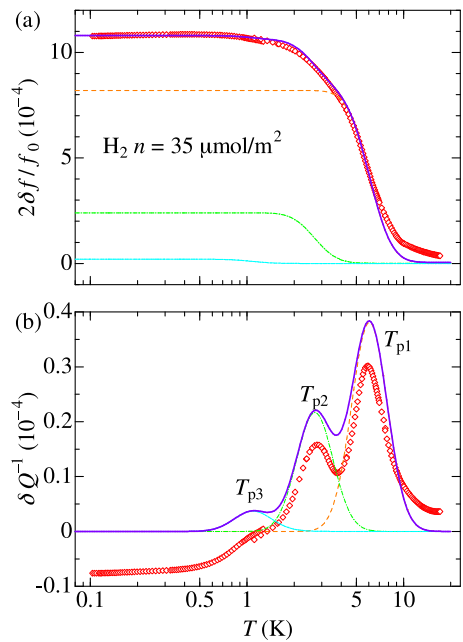}
 \caption{A fitting result for (a) $2\delta f/f_0$ and (b) $\delta Q^{-1}$ with three peaks. The orange, green and blue curves are the fittings for each dissipation peaks using Eq. (\ref{eq:zi}). The purple curve is the sum represented by Eq. (\ref{eq:Z}).}
 \label{fig:Fig3}
\end{figure}

We extract the activation energy by 
fitting the multiple elastic anomalies to the following model function. 
Every single elastic anomaly is described by the dynamic response function \cite{Makiuchi2018},
\begin{equation}
 z_i = \left(\frac{\delta G}{G_0}\right)_i \left[1 - \int_0^\infty \frac{F_i(E)}{1-i\omega \tau_{0i}\exp(E/k_\mathrm{B}T)} dE \right], \label{eq:zi}
\end{equation}
where $\left(\delta G/G_0\right)_i$ is the relaxed shear modulus divided by the shear modulus of the torsion rod, $F_i(E)$ gives the distribution of the activation energy, and $i$ denotes the index of the elastic anomaly.
The normalized frequency shift and excess dissipation are the real and imaginary part of $z_i$, respectively.
We assume $F_i(E) = (1/\sqrt{2\pi}\sigma_i E)\exp(-[\ln(E/\Delta_i)]^2 /2\sigma_i^2)$ (the lognormal distribution) as the best choice \cite{Makiuchi2018,Makiuchi2019}, where $\Delta_i$ is the median of the activation energy and $\sigma$ a dimensionless parameter.
For the multiple elastic anomalies, the total dynamic response function is
\begin{equation}
 Z = \sum_{i=1}^N z_i, \label{eq:Z}
\end{equation}
where $N = 1$, 2, or 3 is the number of the anomaly.
A result of fitting, $2\delta f/f_0 = \mathrm{Re}(Z)$ and $\delta Q^{-1} = \mathrm{Im}(Z)$ with $N=3$, is shown in Fig. \ref{fig:Fig3}.
The model reproduces the step-by-step increase of $2\delta f/f_0$ and multiple peaks of $\delta Q^{-1}$.
The dissipation from the model agrees with the experimental result, except that $\delta Q^{-1}$ of the experiment has some additional $T$-dependent background as in the neon case \cite{Makiuchi2019}.

The ratio of $2\delta f(T\ll T_\mathrm{p})/f_0$ to $\delta Q^{-1}(T_\mathrm{p})$ is essential for determining $\tau_{0i}$. 
 If the ratio is 0.5, $F_i(E)$ is the delta function (no distribution) and $\tau_{0i} = \omega^{-1}\exp (k_\mathrm{B}T_{\mathrm{p}i}/\Delta_i)$.
 If the ratio is smaller than 0.5, as in the present results, 
$\tau_{0i}$ may be smaller in the order of magnitude.
We obtained $\tau_{01}$, $\tau_{02} \sim 10^{-12}$--$10^{-30}$ s for $T_\mathrm{p1}$ and $T_\mathrm{p2}$, which are very small similarly to the neon case, and $\tau_{03}=1\times 10^{-9}$ s which is almost equal to the value in the helium case. 
The parameter $\sigma_i$ was always about 0.3. 
The activation energy $\Delta_i$ was strongly dependent on $T_{\mathrm{p}i}$. 
The fitting parameters at the highest coverages ($n_\mathrm{f}$ for $T_\mathrm{p1}$ and $T_\mathrm{p2}$, and $\sim 40\ \mathrm{\mu mol/m^2}$ for $T_\mathrm{p3}$) 
are shown in Table \ref{tb:ActivationEnergy}.

\begin{table}
 \caption{
 The dissipation peak temperatures and the activation energies at the highest coverages in units of K.
 }
 \label{tb:ActivationEnergy}
 \begin{ruledtabular}
  \begin{tabular}{c|ccc|ccc}
   & $T_\mathrm{p1}$ & $T_{\mathrm{p}2}$ & $T_{\mathrm{p}3}$ & $\Delta_1/k_\mathrm{B}$ & $\Delta_2/k_\mathrm{B}$ & $\Delta_3/k_\mathrm{B}$  \\ \hline
   H$_2$ & 6.0 & 2.8 & 1.1 & 340 & 72 & 13 \\
   HD    & 7.0 & 4.1 & 1.3 & 420 & 180 & 15 \\
   D$_2$ & 7.2 & 4.7 & 2.6 & 420 & 260 & 30 \\ 
  \end{tabular}
 \end{ruledtabular}
\end{table}

We discuss the values of the activation energies.
The activation energies of the classical diffusion in bulk solids were calculated to be $(E_\mathrm{v}+E_\mathrm{b})/k_\mathrm{B}=$ 197 K (H$_2$) and 290 K (D$_2$) \cite{Ebner1972}.
Our results of $\Delta_1/k_\mathrm{B}$'s give higher values.
The activation energies for the quantum tunneling in bulk, i.e., energy to create vacancies, were $E_\mathrm{v}/k_\mathrm{B}=$ 112 K (H$_2$) and 132 K (D$_2$) \cite{Ebner1972}.
Our results of $\Delta_2/k_\mathrm{B}$ are larger in D$_2$ but smaller in H$_2$.
The larger values of $\Delta_i$ are attributed to the molecular confinement in nanopores.
The activation energies in the first and second layers on the substrate can be larger than that of bulk because these layers are strongly bound and compressed.
In the case of H$_2$, the larger zero-point fluctuation may reduce $\Delta_2$.

For the surface diffusion, $\Delta_3$'s roughly agree with activation energies in previous studies;
$E_\mathrm{s}/k_\mathrm{B}=(19\pm 2)$ (H$_2$) \cite{Sukhatme1996},
$(43\pm 7)$ K (H$_2$) \cite{Bloss2000},
$(16.1\pm 0.5)$ K (HD) \cite{Maruyama1993}, and
$(38\pm 4)$ K (D$_2$) \cite{Sukhatme1996}.
The differences in quantity are probably due to the differences of the film thickness and the substrate.
Our results of $\Delta_{i}$ give relatively small values.

$T_\mathrm{p3}$'s in Fig. \ref{fig:Fig2} do not decrease below 1 K. 
This means that hydrogen films do not undergo a QPT, contrary to helium \cite{Makiuchi2018}. 
However, the concave curvatures of $T_\mathrm{p3}$ of hydrogen and $T_\mathrm{p}$ of helium are very similar in $n_1<n<1.8n_1$.
Moreover, the ratio $\Delta_3/k_\mathrm{B}T_\mathrm{p3}$ is 12 for H$_2$, HD, and D$_2$ films.  
It agrees well with $\Delta/k_\mathrm{B}T_\mathrm{p}=13$ of $^4$He and $^3$He films \cite{Makiuchi2019}.
These similarities between hydrogen and helium strongly suggest that the third elastic anomaly in hydrogen films is also originated from a quantum many-body ground state, such as Mott insulator or Mott glass\cite{Giamarchi2001}. 
$\Delta_3$ is regarded as a Mott gap, and at finite temperatures the surface molecules are excited to spatially extended states. 
The surface layers are \textit{on the verge of a QPT to} (\textit{super})\textit{fluid state}. 

Finally, we propose that, if the surface molecules are excited at low temperatures by artificial means, e.g. emission of phonons with frequency $\omega > \Delta_3/\hbar$, they can be macroscopically condensed and exhibit superfluidity in a non-equilibrium sense. 
Study of the relaxation rate from excited to ground states is needed to examine the feasibility of non-equilibrium superfluidity.

To conclude, we discovered multiple elastic anomalies in all hydrogen isotope films, in which the elastic constant has multiple steps accompanied with dissipation peaks. 
The multiple anomalies correspond to the ``freezing'' of different diffusion mechanisms: the classical thermal diffusion, the quantum tunneling, and the surface diffusion of molecules.
The uppermost surface of the films is in a quantum many-body ground state, which is on the verge of a quantum phase transition to (super)fluid state. 
Our finding in this work implies new approaches to the realization of superfluidity in hydrogen films. 

\begin{acknowledgments}
 This work has been supported by JSPS KAKENHI Grant No. JP17H02925, JP17K18762, and JP19K21856. 
 T.M. was supported by Grant-in-Aid for JSPS Research Fellow JP18J13209.
\end{acknowledgments}





\begin{thebibliography}{27}
\bibitem [{sam()}]{samplegas}%
  \BibitemOpen
  \href@noop {} {}\bibinfo {note} {The sample gas purity reported by the
  supplier was $99.99999$\%, $97$\%, and $96$\% for H$_2$, HD, and D$_2$,
  respectively.}\BibitemShut {Stop}%
\end{thebibliography}
%
\providecommand{\noopsort}[1]{}\providecommand{\singleletter}[1]{#1}%

\end{document}